\documentclass[utf8]{FrontiersinHarvard} 

\usepackage{url,hyperref,lineno,microtype,subcaption}
\usepackage[onehalfspacing]{setspace}
\usepackage{aas_macros}

\def\keyFont{\fontsize{8}{11}\helveticabold }
\def\firstAuthorLast{Corsi {et~al.}} 
\def\Authors{Alessandra Corsi\,$^{1,*}$, Lisa Barsotti$^{2}$, Emanuele Berti$^{3}$, 
Matthew Evans$^{2}$, Ish Gupta$^4$, Konstantinos Kritos$^{3}$, Kevin Kuns$^{2}$, Alexander\,H.\,Nitz$^{5}$, Benjamin\,J.\,Owen\,$^{1}$, Binod Rajbhandari$^6$, Jocelyn Read$^{7}$, Bangalore\,S.\,Sathyaprakash$^4$, David\,H.\,Shoemaker$^{2}$, Joshua\,R.\,Smith$^{7}$, Salvatore Vitale$^{2}$}
\def\Address{$^{1}$Department of Physics and Astronomy, Texas Tech University, Lubbock, TX 79409, USA \\
$^2$ LIGO Laboratory, Massachusetts Institute of Technology, Cambridge, MA 02139, USA\\
$^3$ Department of Physics and Astronomy, Johns Hopkins University, Baltimore, MD, USA\\
$^4$ Institute for Gravitation and the Cosmos, Department of Physics, Pennsylvania State University, University Park, PA 16802, USA\\
$^5$Department of Physics, Syracuse University, Syracuse, NY 13244, USA\\
$^6$ School of Mathematical Sciences and Center for Computational Relativity and
Gravitation, Rochester Institute of Technology, Rochester, NY 14623, USA\\
$^7$ Nicholas and Lee Begovich Center for Gravitational Wave Physics and Astronomy, California State University Fullerton,
Fullerton CA 92831, USA
}

\def\corrAuthor{Corresponding Author}

\def\corrEmail{alessandra.corsi@ttu.edu}

\begin{document}
\onecolumn
\firstpage{1}

\title[MMA of BHs and NSs]{Multi-messenger Astrophysics of Black Holes and Neutron Stars as Probed by Ground-based Gravitational Wave Detectors: From Present to Future} 

\author[\firstAuthorLast ]{\Authors} 
\address{} 
\correspondance{} 

\extraAuth{}

\maketitle
\begin{abstract}

The ground-based gravitational wave (GW) detectors LIGO and Virgo have enabled the birth of multi-messenger GW astronomy via the detection of GWs from merging stellar-mass black holes (BHs) and neutron stars (NSs). GW170817, the first binary NS merger detected in GWs and all bands of the electromagnetic spectrum, is an outstanding example of the impact that GW discoveries can have on multi-messenger astronomy. Yet, GW170817 is only one of the many and varied multi-messenger sources that can be unveiled using ground-based GW detectors. In this contribution, we summarize key open questions in the astrophysics of stellar-mass BHs and NSs that can be answered using current and future-generation ground-based GW detectors, and highlight the potential for new multi-messenger discoveries ahead.

\tiny
 \keyFont{ \section{Keywords:} Gravitational Waves; Time-domain Astronomy; Multi-messenger Astrophysics; GW170817; LIGO and VIRGO} 
\end{abstract}

\section{Introduction}
\begin{figure}
\begin{center}
\includegraphics[width=0.8\textwidth]{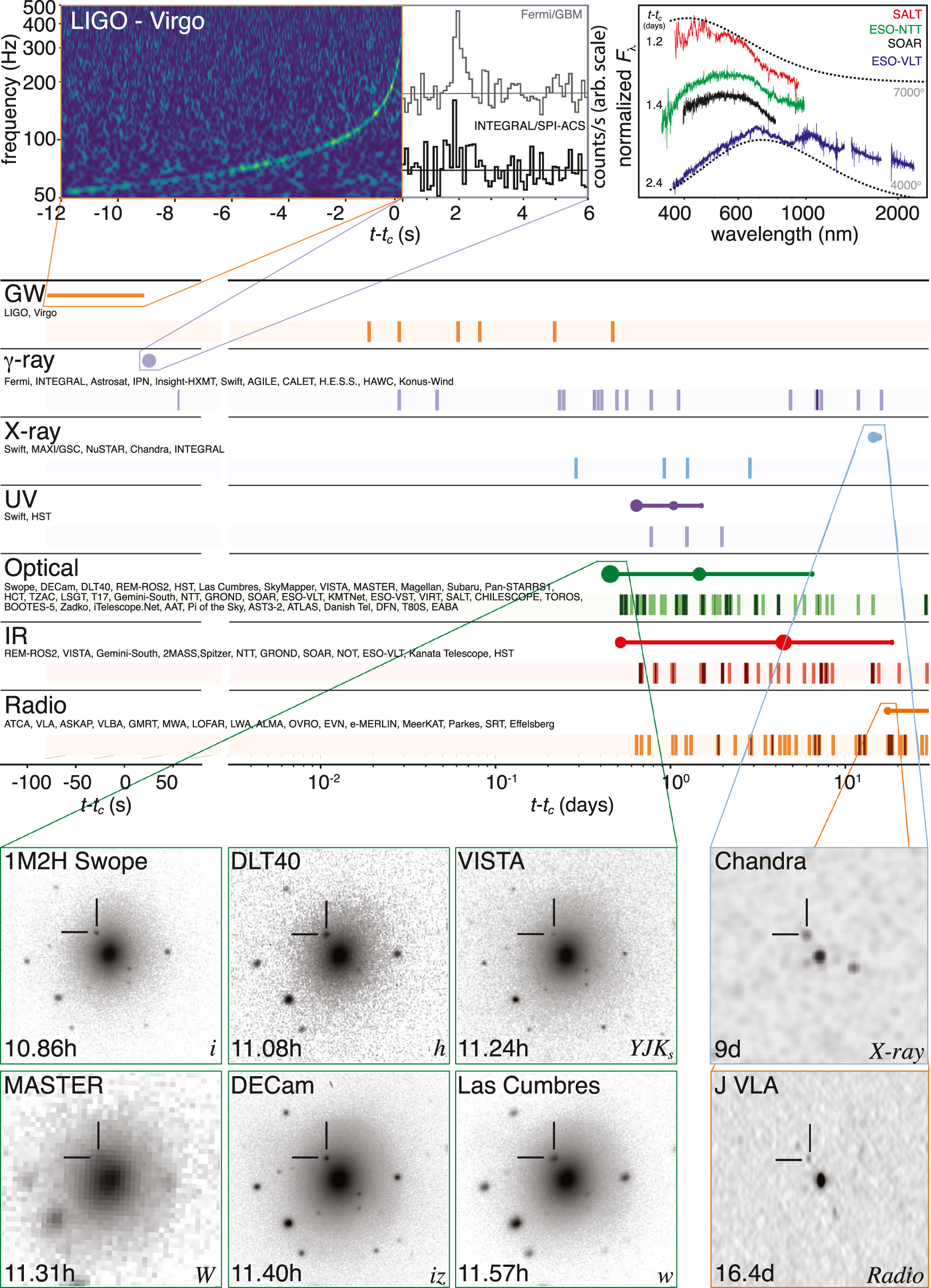}
\end{center}
\caption{Figure reproduced from \citet{2017ApJ...848L..12A}. Timeline of the discovery of GW170817, its associated GRB\,170817a, and its associated kilonova SSS17a/AT 2017gfo. The follow-up observations are shown by messenger and wavelength relative to the time of the GW event. The shaded dashes represent the times when information was reported in a GCN Circular. The names of the relevant instruments, facilities, or observing teams are collected at the beginning of the row. Representative observations in each band are shown as solid circles with their areas approximately scaled by brightness; the solid lines indicate when the source was detectable by at least one telescope. Magnification insets give a picture of the first detections in GWs, and in the gamma-ray, optical, X-ray, and radio bands.}
\label{fig:GW170817}
\end{figure}

The discovery of the binary NS merger GW170817 during the second observing run (O2) of the LIGO \citep{2015CQGra..32g4001L} and Virgo \citep{2015CQGra..32b4001A} GW detectors kicked off a new era in multi-messenger astrophysics (MMA; Figures \ref{fig:GW170817}-\ref{fig:observatories}). In addition to marking the first direct detection of a GW chirp from a binary NS merger \citep{2017PhRvL.119p1101A}, GW170817 also represents the first astrophysical event to be observed with GWs and a completely independent messenger, namely, electromagnetic waves. Indeed, GW170817 was the first direct association of a NS-NS merger with a short gamma-ray burst (GRB), an IR-optical-UV kilonova, and an electromagnetic afterglow observed from radio to X-rays \citep[see][and references therein]{2017ApJ...848L..12A}. 

The rich multi-messenger data collected for GW170817 (Figure \ref{fig:GW170817}), together with detailed modeling and simulations, have painted the most detailed picture yet of a binary NS merger, impacting a variety of fields beyond gravitational physics and including nuclear physics \citep[e.g.,][]{2017ApJ...850L..34B,2017Natur.551...80K,2017ApJ...850L..19M,2018PhRvL.121p1101A,2018PhRvL.120q2703A,2018PhRvL.121i1102D,2018ApJ...855...99C,2018PhRvL.120z1103M,2018ApJ...852L..29R,2018ApJ...852L..25R,2020NatAs...4..625C}, relativistic astrophysics \citep[e.g.,][]{2017PhRvD..96l3012S,2018PhRvL.120x1103L,2018PhRvD..97b1501R,2021ApJ...918L...6L}, stellar evolution and population synthesis \citep[e.g.,][]{2013ApJ...779...72D,2018MNRAS.481.1908K,2018MNRAS.481.4009V}, and cosmology \citep[e.g.,][]{2017Natur.551...85A,2017PhRvL.119y1301B,2017PhRvL.119y1302C,2017PhRvL.119y1304E,2017PhRvL.119y1303S,2018Natur.562..545C}. 

As of today, the LIGO and Virgo detectors have reported highly-significant discoveries of $\sim 100$ compact binary coalescences \citep{2023PhRvX..13d1039A}. The detections are dominated by binary BH mergers. Two highly-significant NS-NS mergers \citep[GW170817 and GW190425;][]{2017PhRvL.119p1101A,2020ApJ...892L...3A} and a few BH-NS merger candidates have been identified \citep{2021ApJ...915L...5A}, but GW170817 remains the only GW event with a secure electromagnetic counterpart association. While revolutionizing the field of GW-MMA, the discovery of GW170817 highlighted many open questions that remain to be answered. To this end, the LIGO and Virgo collaborations have developed plans for further improvements in sensitivity of these detectors that will fully exploit what is possible at these existing facilities \citep[hereafter, post-O5 or A\# era, Figure \ref{fig:sensitivity};][]{2018LRR....21....3A,post-O5}.  Several new frontiers in MMA will also come on the horizon with these envisioned sensitivity upgrades for the LIGO detectors \citep[which include an expanded network with LIGO India---hereafter LIGO  Aundha---expected to be operational starting in the early 2030;][]{LIGOIndia}. However, it is likely that the full discovery potential of MMA will be realized only with next generation ground-based GW detectors such as Cosmic Explorer (hereafter, CE) and the Einstein Telescope (hereafter, ET), envisioned to become operational in the 2030s and requiring new facilities and longer interferometer arms \citep[Figures \ref{fig:observatories}-\ref{fig:sensitivity};][]{2023JCAP...07..068B,2023arXiv230613745E}. Here, we review the major open questions in the field of MMA as enabled by ground-based GW detectors (Sections \ref{sec:science}--\ref{sec:frontiers}), and briefly discuss the short-to-long term potential of this field  (Section \ref{sec:discussion}). 

We stress that, while this work highlights topics in MMA for which observations of GWs and light are critical, the field of MMA is broader and includes messengers such as cosmic rays and neutrinos \citep[e.g.,][and references therein]{P5report}. Here, we mention these other probes only briefly. We also stress that our discussion is centered on the science enabled by ground-based GW detectors  operating in the few Hz to few kHz GW frequency regime. However, the GW spectrum is much broader, and fundamental contributions to its exploration are being provided by Pulsar Timing Arrays \citep{1979ApJ...234.1100D,2023ApJ...951L...8A,2023A&A...678A..50E}, and will be provided in the future by space-based instruments such as LISA \citep{2023LRR....26....2A} and DECIGO \citep{2011CQGra..28i4011K}. 

\section{MMA of compact binary mergers: Key open questions}
\label{sec:science}

\subsection{Diversity of NS-NS/BH-NS mergers and r-process yields}
\label{sec:diversity}
GW170817 remains so far the only event seen in both GWs and electromagnetic emission. An associated GRB (170817A) was detected about 2\,s after the merger by the \textit{Fermi}/GBM and Integral satellites \citep[Figure \ref{fig:GW170817};][]{2017ApJ...848L..13A,2017ApJ...848L..15S}. About 11\,hr after the GW detection, an optical counterpart was identified by the Swope Supernova Team \citep[Figure \ref{fig:GW170817};][]{2017Sci...358.1556C}. Via extensive multi-wavelength observations carried by several teams, this counterpart was recognized to be a kilonova---a quasi-thermal fast-fading transient associated with r-process nucleosynthesis occurring in the neutron-rich debris created by the merger itself \citep{2017ApJ...848L..19C,2017ApJ...848L..17C,2017Sci...358.1570D,2017Sci...358.1565E,2017Sci...358.1559K,2017ApJ...848L..18N,2017Natur.551...67P,2017Natur.551...75S,2017ApJ...848L..16S,2017ApJ...848L..27T,2017ApJ...848L..24V,2017ApJ...851L..21V}. The kilonova detection also enabled the arcsec localization of GW170817, and hence the identification of its host galaxy and measurement of its redshift \citep{2017ApJ...849L..16I,2017ApJ...848L..31H,2017ApJ...848L..28L,2017ApJ...849L..34P,2017ApJ...848L..30P}. Located only $\approx 40$\,Mpc away, GW170817 is the closest short GRB with known redshift identified as of today. As the radio-to-X-ray follow-up observations of the GW170817/GRB\,170817A afterglow revealed, GW170817 also brought the first ever direct detection of a relativistic jet observed off-axis (Figure \ref{fig:GW170817}, JVLA and \textit{Chandra} insets), and proved that relativistic jets are much more complex than typically assumed for cosmological short GRBs \citep[for which the on-axis view prevents a detailed study of the jet structure;][]{2017ApJ...848L..21A,2017ApJ...848L..25H,2017Sci...358.1579H,2017ApJ...848L..20M,2017Natur.551...71T,2018Natur.554..207M,2018ApJ...856L..18M,2018Natur.561..355M}. 

\begin{figure}
\begin{center}
\hbox{
\includegraphics[height=3.5cm]{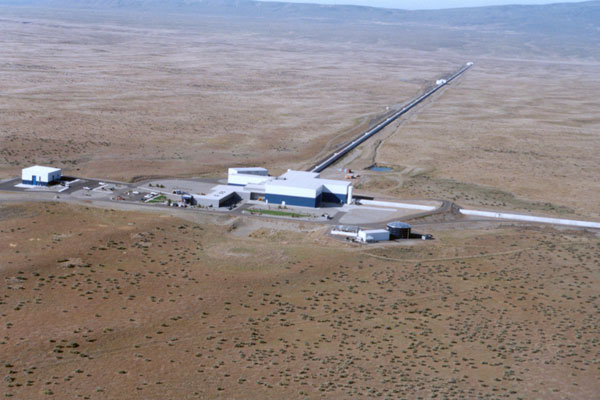}
\includegraphics[height=3.5cm]{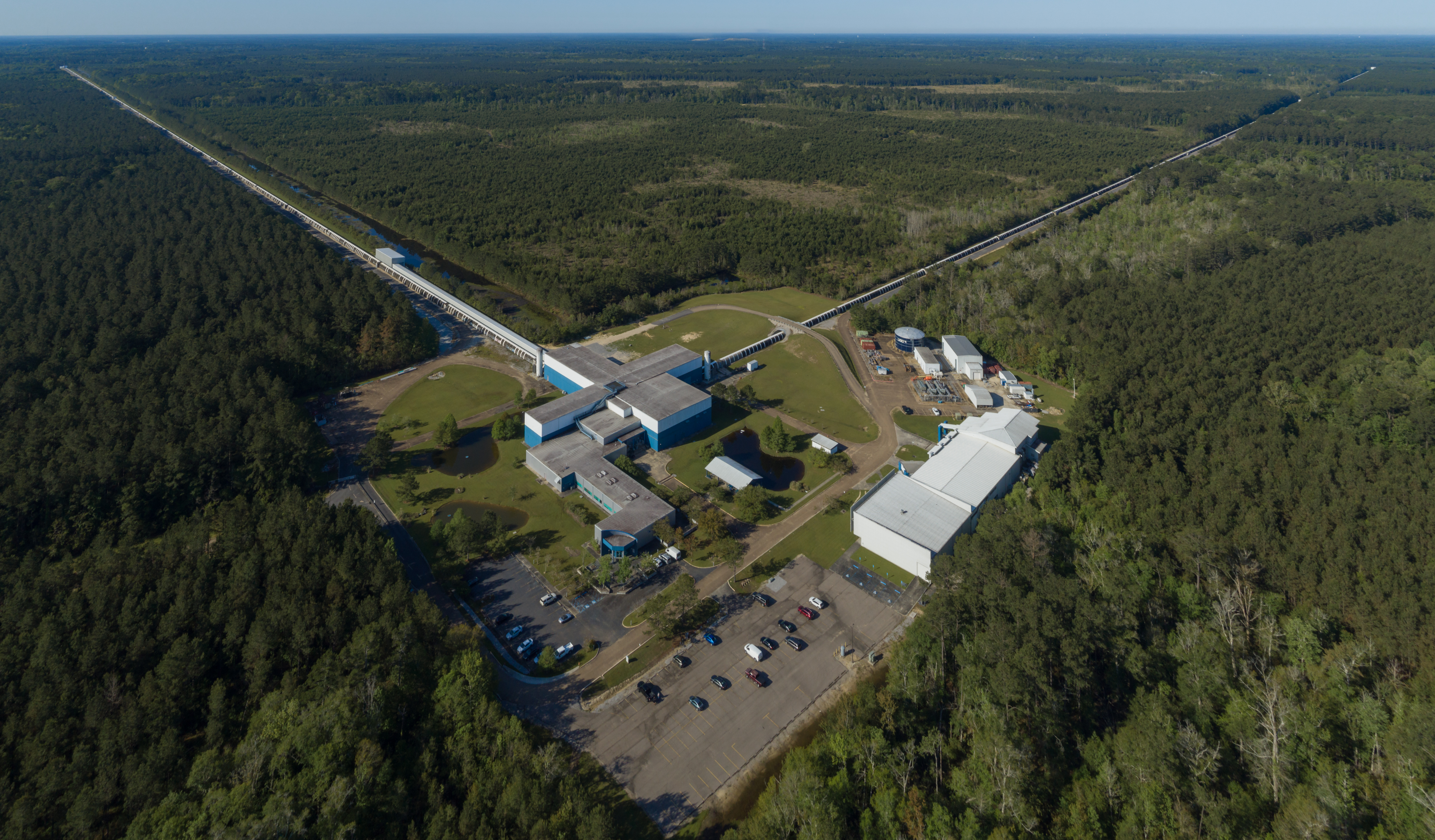}
\includegraphics[height=3.5cm]{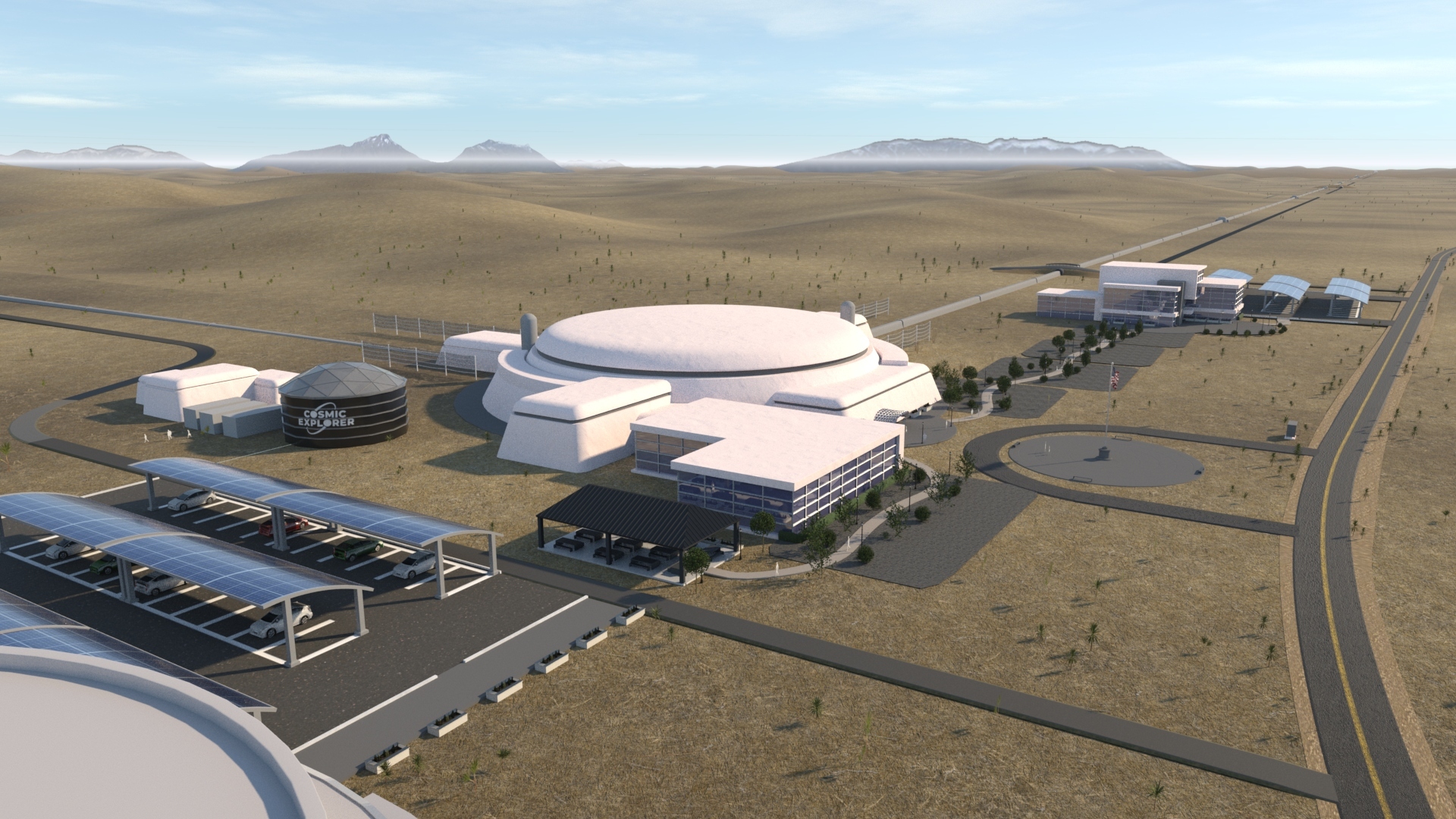}
}
\end{center}
\caption{Aerial views of the LIGO Hanford (left) and Livingston (center) observatories \citep[credits: Caltech/MIT/LIGO Lab;][]{2015CQGra..32g4001L}. We also show an artist's impression of a Cosmic Explorer (CE) observatory \citep[credits: Angela Nguyen, Virginia Kitchen, Eddie Anaya, California State University Fullerton;][]{2023arXiv230613745E}.}
\label{fig:observatories}
\end{figure}

As the sensitivity of the LIGO detectors continues to improve steadily compared to the O2 run (Figure \ref{fig:sensitivity}), one of the biggest priorities in the field of MMA is the collection of a larger sample of GW170817-like multi-messenger detections: going from 1 to $\sim 10$ (nearby) events localized by GW detectors to less than $100$\,deg$^2$ by the post-O5 / A\# era is a must \citep[][see also Section \ref{sec:discussion}]{2022ApJ...924...54P,2018LRR....21....3A}. Increasing the sample of nearby, extensively monitored events is key to answering some fundamental questions left open by GW170817 such as, are NS-NS mergers the only site or one of many sites of r-process nucleosynthesis; are the heaviest of the heavy elements synthesized in those mergers; does the yield of various heavy elements match the Solar abundance \citep[e.g.,][and references therein]{1989Natur.340..126E,2013ApJ...773...78B,2019LRR....23....1M,2023A&ARv..31....1A,2023MNRAS.520.2829S}.  More generally, nearby multi-messenger detections are critical to understanding what is the possible zoo of electromagnetic counterparts of NS-NS and BH-NS systems (blue versus red kilonovae, choked versus successful and structured versus top-hat jets), and what is the range of circum-burst medium densities in relation to the properties of the host galaxies \citep[e.g.,][]{2002AJ....123.1111B,2013ApJ...776...18F,2016ApJ...829..110B,2016ApJ...831..190H,2016MNRAS.460.3255R,2017PhRvD..96l4005B,2019ApJ...880L..15M,2019ApJ...881...89L,2020PhR...886....1N,2021JPlPh..87a8402A,2021ApJ...922..269R,2022ApJ...940...56F,2022MNRAS.516.4760C,2022MNRAS.517.1640G,2022MNRAS.512.2654P,2023MNRAS.526.4585G,2023arXiv231016894C,2023ApJ...944..220N}. 

Ultimately, a diverse sample of multi-messenger detections of nearby and well-localized NS-NS and BH-NS systems will enable us to map the properties of the progenitors as probed by GWs \citep[especially in terms of total mass, mass ratio, and Equation of State, hereafter, EoS;][and references therein]{2018PhRvL.121p1101A,2019PhRvX...9a1001A}, to the properties of their merger ejecta and of the circum-merger environment as probed by electromagnetic observations \citep[][and references therein]{2019ApJ...880L..15M}. Joint multi-messenger analysis will then shed light on the physical processes that determine such mapping \citep[e.g.,][and refrences therein]{2018ApJ...869..130R}.

\subsection{Short GRB jets and central engines}
\label{sec:jets}
\begin{figure}
\begin{center}
\includegraphics[width=0.7\textwidth]{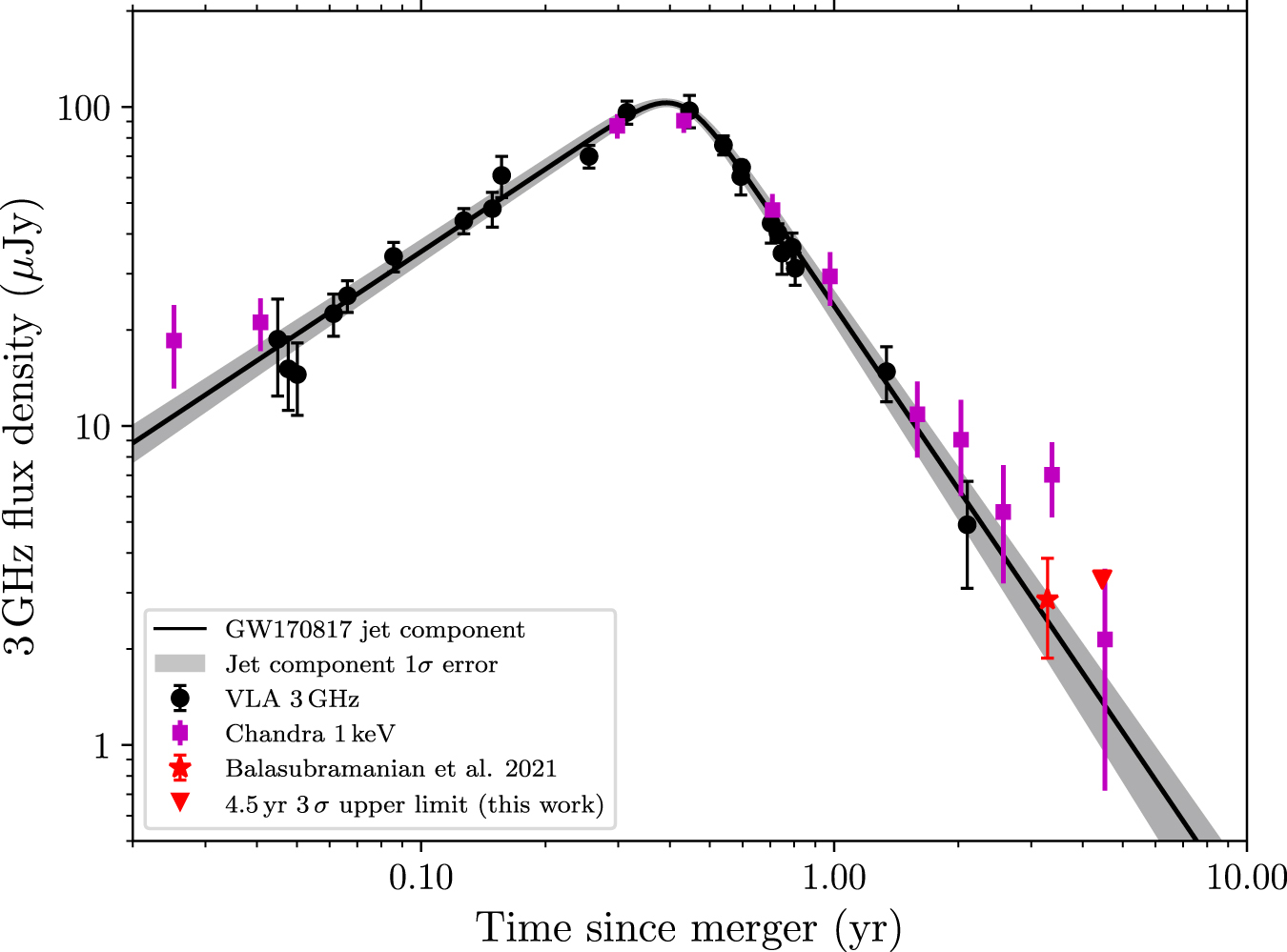}
\end{center}
\caption{Figure reproduced from \citet{2022ApJ...938...12B}. 3\,GHz radio light curve of GW170817 (black dots, red star and red triangle) plus extrapolation of the X-ray observations to the radio band (purple squares), together with the best fit model and corresponding error (black line and gray shaded area) representing the emission from the relativistic jet. }
\label{fig:radio}
\end{figure}

The association of GW170817 with a GRB and an off-axis radio-to-X-ray afterglow (Section \ref{sec:diversity}; Figure \ref{fig:radio}) has demonstrated how GW observations can open the way to directly linking GRB progenitors to their relativistic jets. However, we are still far from fully understanding the physics behind the workings of GRB central engines and their jets, especially in terms of emission processes, jet composition and structure, and the role of magnetic fields.

\begin{figure}
\begin{center}
\includegraphics[width=0.8\textwidth]{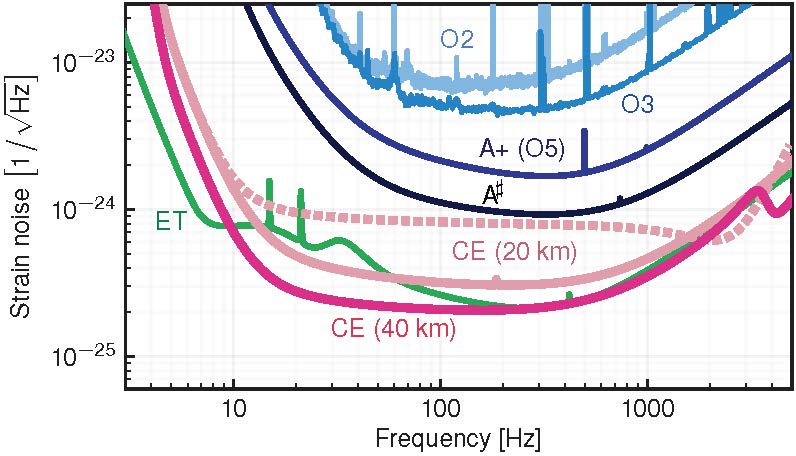}
\end{center}
\caption{Figure adapted from \citet{2023arXiv230613745E}. Measured sensitivity of LIGO in its second (O2) and third (O3) observing runs, and estimated sensitivities of LIGO A+ \citep[also referred to as LIGO O5 sensitivity;][]{2018LRR....21....3A}, LIGO A\# \citep{post-O5}, ET \citep[][]{2023JCAP...07..068B}, and the 20\,km and 40\,km CE detectors \citep{2023arXiv230613745E}. We note that by reconfiguring several smaller optics, the 20\,km detector could be operated either in a broad-band mode (solid) or a kilohertz-focused mode (dotted). }
\label{fig:sensitivity}
\end{figure}

Because in a compact binary merger the amplitude of the emitted GWs depends mildly on the orientation of the binary, GW detections can enable the study of off-axis GRB jets that may otherwise go undetected and/or unrecognized as off-axis events via electromagnetic observations alone \citep{2017MNRAS.471.1652L,2018MNRAS.481.1597G,2019MNRAS.485.4150B,2020MNRAS.492.5011D,2020MNRAS.492.4283M,2020ApJ...902...82S,2021ApJ...908...63G,2021MNRAS.500.1708R,2023ApJ...948..125E,2024MNRAS.527.8068G}. This is key to shedding light on the jet structures that, in turn, are determined by complex processes involving the GRB central engines (that power the jet itself), and the interaction of the jets with the neutron-rich debris surrounding the merger sites \citep[e.g., ][and references therein]{2002MNRAS.332..945R,2005A&A...436..273A,2011ApJ...740..100B,2018MNRAS.478..407N,2018PhRvL.120x1103L,2019ApJ...881...89L,2021MNRAS.500.3511G,2022arXiv220611088S,2023MNRAS.519.4454G,2023MNRAS.524..260P}. 
As demonstrated by $\sim 50$ years of GRB observations, the structure of relativistic jets is largely masked in high-luminosity cosmological GRBs, whose electromagnetic emission is dominated by fast jet cores observed on-axis. In fact, the prompt $\gamma$-ray emission of the off-axis GRB\,170817A was energetically weaker by about three orders of magnitude than the weakest cosmological short GRB \citep{2015ApJ...815..102F}. Its afterglow showed a behavior substantially different from the power-law-decaying afterglows of cosmological GRBs, with a delayed onset and a rising light curve observed from radio to X-rays \citep[Figure \ref{fig:radio}][]{2017ApJ...848L..21A,2017ApJ...848L..25H,2017Sci...358.1579H,2017ApJ...848L..20M,2017Natur.551...71T,2018Natur.554..207M,2018ApJ...856L..18M,2018Natur.561..355M,2021ApJ...922..154M,2022ApJ...938...12B}. While extensive multi-band observations and detailed modeling have allowed us to link these unusual properties of GRB\,170817A with a structured jet observed off-axis \citep{2018PhRvL.120x1103L}, significant uncertainties remain.  Specifically,  the polar profile (distribution of energy as a function of polar angle) of the GW170817 outflow remains highly debated, with analytical functions including Gaussian, power-law, and exponential profiles, as well as numerically-simulated profiles, all providing plausible fits to the data. 
In the the radio band, future observations of off-axis GRB light curves combined with polarization measurements and Very-Long Baseline Interferometry (VLBI) can help shed light on both the jet structure and the largely unknown structure of magnetic fields within shocked ejecta \citep[e.g.,][and references therein]{1999MNRAS.309L...7G,1999ApJ...524L..43S,2018ApJ...861L..10C,2018MNRAS.478.4128G,2018MNRAS.481.2711G,2018Natur.561..355M,2019Sci...363..968G,2020MNRAS.491.5815G}. 

The origin of the $\gamma$-rays in GRB170817a remains equally debated: while the structured outflow model can explain why a GRB was detected even if off-axis \citep{2017MNRAS.471.1652L}, a mildly relativistic shock breakout of a cocoon from the merger's ejecta is also possible \citep{2018MNRAS.479..588G}. Future multi-messenger observations of off-axis GRBs \citep[including potential coincident detections between GW signals and sub-threshold GRBs;][]{2018ApJ...862..152K,2019ApJ...878L..17M,2020ApJ...900...35T,2023arXiv230813666F}, will greatly help settle these debates \citep{2020FrASS...7...78L,2022MNRAS.515..555B,2022Galax..10...38B}.  

\begin{figure}
\begin{center}
\includegraphics[width=\textwidth]{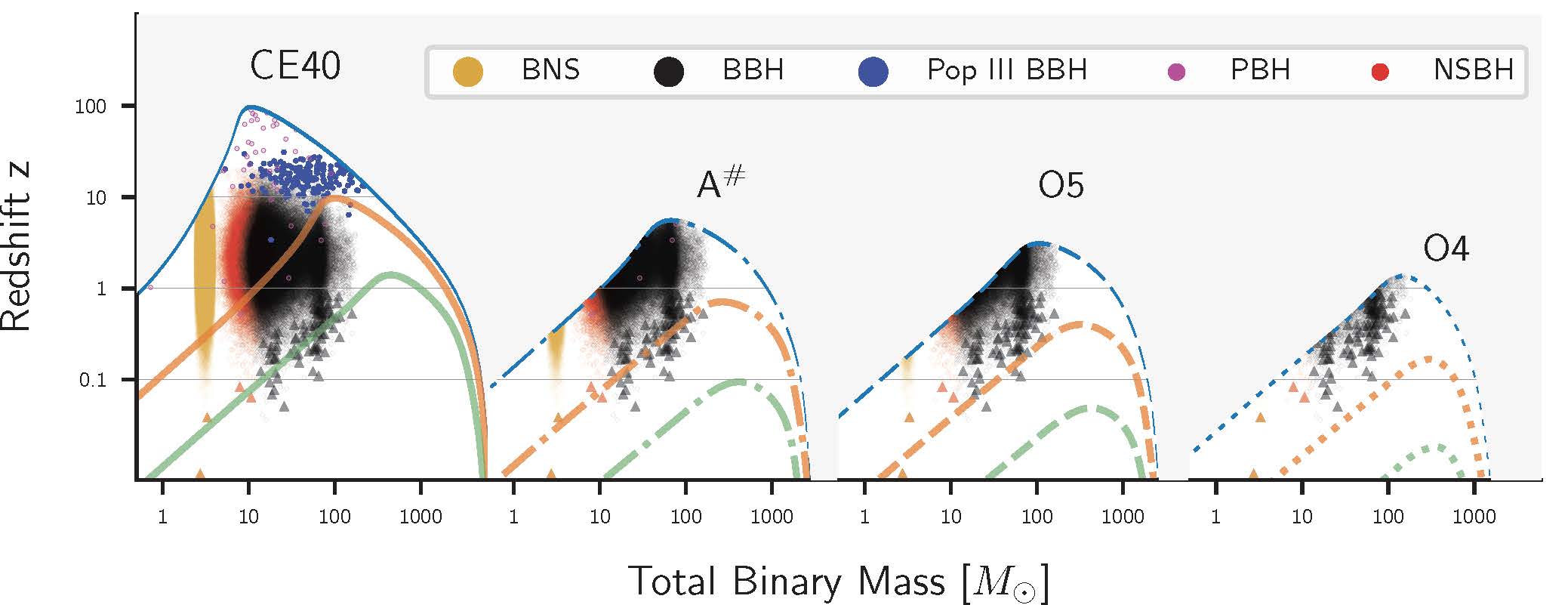}
\caption{Figure adapted from \citet{2023arXiv230613745E}. The reach of current and future ground-based GW detectors for compact binary mergers (NS-NS mergers in gold; BH-NS mergers in red; and BH-BH mergers in black; see Section \ref{sec:populations}) is represented as a function of total binary mass and redshift at various signal-to-noise ratio (SNR) thresholds (blue lines for SNR 8; orange lines for SNR 100; and green lines for SNR 1000). The population of observed
compact-object binaries is plotted with small triangles. We use dotted lines for LIGO at its O4 sensitivity; dashed lines for LIGO at its projected O5 sensitivity, also referred to as LIGO A+ \citep{2018LRR....21....3A}; and dash-dotted lines for LIGO at its projected post-O5 A\# sensitivity \citep[the ultimate performance of current LIGO detectors envisioned for the post-O5 era;][]{post-O5}. CE40 \citep[][]{2023arXiv230613745E}, a next generation GW detector concept, can expand the cosmic horizon of NS-NS mergers, and enable observations of new populations including mergers from Population III BHs (blue dots), and speculative primordial BHs
(magenta dots).}
\label{fig:reach}
\end{center}
\end{figure}

While the LIGO-Virgo-KAGRA detectors (Figure \ref{fig:observatories}, left and central panel) continue to improve their sensitivity to GWs from GRBs \citep{2021ApJ...915...86A,2022ApJ...928..186A}, these searches will undergo a leap forward when next-generation GW detector such as CE and ET,  with $\approx 10\times$ the sensitivity of the current LIGO detectors (Figure \ref{fig:observatories}, right panel, and Figure \ref{fig:sensitivity}), will probe the population of NS-NS mergers up to the star formation peak \citep[and beyond for BH-BH mergers, Figure \ref{fig:reach} and ][]{2023JCAP...07..068B,2023arXiv230613745E,2023arXiv230710421G}. With these next generation detectors, we can expect each short GRB observed by satellites such as \textit{Fermi} \citep{2022hxga.book...29T} and \textit{Swift} \citep{2004ApJ...611.1005G} to have a counterpart in GWs \citep{2022AA...665A..97}. The direct mapping of GRBs to their progenitors---something inaccessible to electromagnetic observations alone---is key to shedding light on the conditions that enable the launch of successful relativistic jets, especially in relation to the properties of the progenitors \citep[including whether BH-BH mergers make GRBs;][]{2016ApJ...819L..21L,2016ApJ...826L...6C,2017MNRAS.470L..92D,2018MNRAS.477.4228P,2019ApJ...875...49P,2019ApJ...882...53V,2023ApJ...942...99G} and the nature of the central engines \citep[BHs versus long- or short-lived NSs;][see also Section \ref{sec:postmerger}]{2012MNRAS.419.1537B,2013ApJ...771L..26G,2013PhRvL.111m1101B,2013ApJ...762L..18G,2020ApJ...901L..37M}. Systematic measurements of the delay times between GW mergers and GRBs, in addition to providing stringent fundamental physics tests, will further our understanding of the GRB jet launching mechanisms, of the physics of the jet breakouts from the surrounding medium, and of the dissipation and radiation mechanisms as related to the unknown composition of jets \citep{2017ApJ...850L..24G,2018PhRvD..97h3013S,2019FrPhy..1464402Z,2020FrASS...7...78L}. 

Probing directly and systematically the progenitor of short GRBs observed in $\gamma$-rays will also shed light on whether the phenomenological classification of GRBs in short/hard and long/soft as related to two different classes of progenitors (compact binary mergers and collapsars, respectively) holds in all cases. In fact, this classification scheme has been challenged by observations of long GRBs associated with kilonovae or lacking supernova counterparts to very deep limits, and short GRBs showing potential supernova bumps in their light curves \citep{2006Natur.444.1050D,2006Natur.444.1047F,2021NatAs...5..917A,2022Natur.612..223R,2022ApJ...932....1R,2022Natur.612..228T,2022Natur.612..232Y,2023NatAs...7...67G,2023ApJ...947...55B}. In the future, deep GW observations of these peculiar GRBs will provide the definitive word on the nature of their progenitors and likely settle current classification debates \citep[][]{2023ApJ...949L..22D}. 

\begin{figure}
\begin{center}
\includegraphics[width=\textwidth]{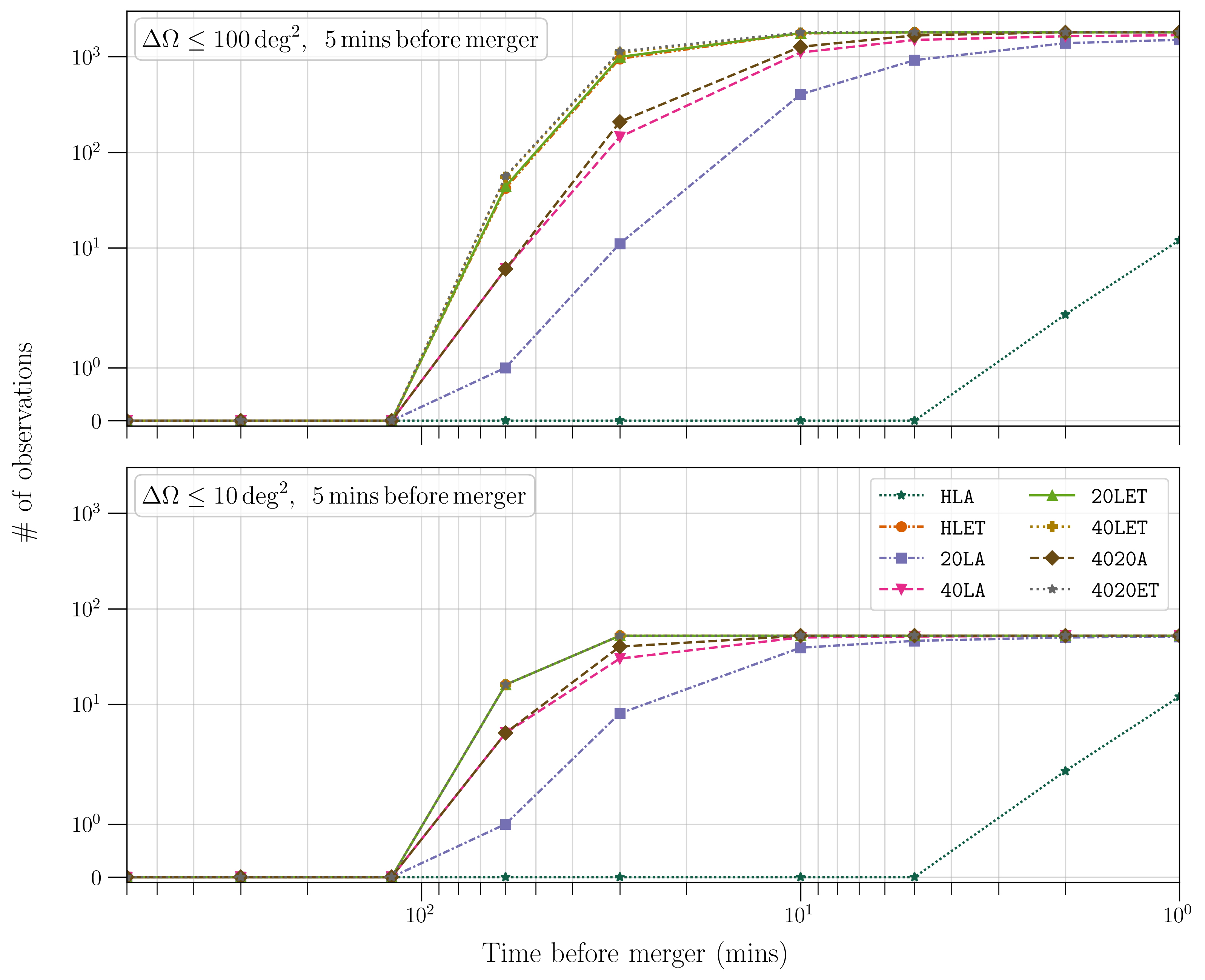}
\caption{This Figure is based on the simulations presented in \citet{2023arXiv230710421G}, for GW detector networks containing zero to three next-generation observatories. The HLA network contains the two current LIGO detectors (Hanford and Livingston) operating at the upgraded A\# sensitivity (Figure \ref{fig:sensitivity}), plus the LIGO Aundha at A\# sensitivity. The 20LA and 40LA networks represent configurations with a single 20\,km-long arms CE detector operating in the context of an upgraded (A\# sensitivity) LIGO network with locations in Livingston and Aundha. The HLET network is one with a single next generation GW detector (ET) operating together with LIGO Hanford and LIGO Livingston at their upgraded A\# sensitivity. The 4020A network represents the CE reference configuration as described in \citep{2023arXiv230613745E}, with one 40\,km-long and one 20\,km-long next generation detectors plus LIGO Aundha at A\# sensitivity. The 20LET and 40LET networks represent a single CE detector (either 20\,km or 40\,km) operating with LIGO Livingston and the ET. Finally, the  4020ET is the reference CE configuration operating with ET. For these networks, we calculate the signal-to-noise ratio of NS-NS systems at 1, 2, 5, 10, 30, 60, 120, 300, 600\,min before merger (data points) for events that are localized within 100\,deg$^2$ (top) or 10\,deg$^2$ (bottom) in 1\,yr. If the network signal-to-noise ratio is $> 10$ at the considered time before merger, then the binary is included in the count. We assume a local merger rate density of $320$\,Gpc$^{-3}$\,yr$^{-1}$, but note that this rate is subject to large uncertainties \citep[$10-1700$\,Gpc$^{-3}$\,yr$^{-1}$;][]{2023PhRvX..13a1048A}. There are no events satisfying the imposed criteria at $> 120$\,min before the merger given the assumed low-frequency cut-off of 5\,Hz for all the detectors (results could be improved if ET reaches sub-5Hz sensitivity). We also note that all events with $\Delta\Omega\le 10$\,deg$^2$ at 5\,min before merger are located at $z<0.2$; and, all events with $\Delta\Omega\le 100$\,deg$^2$ at 5\,min before merger are located at $z<0.5$. Finally, all events detected 5\,min before merger (with no restrictions on the localization accuracy) lie at $z < 0.9$.}
\label{fig:precursor}
\end{center}
\end{figure}

\subsection{Electromagnetic precursors to compact binary mergers}
\label{sec:precursosrs}
Electromagnetic emission from GW170817 was probed only after the GW merger (starting from about 2\,s after; Figure \ref{fig:GW170817}) with the detection of $\gamma$-rays. Hence, as of today, the pre-merger phase remains unexplored in terms of potential electromagnetic counterparts. As the sensitivity and number of ground-based GW detectors increase, GW observations of an in-spiraling system can provide the advance notice required to capture light from the moments closest to merger \citep[Figure \ref{fig:precursor}; see also][]{2012ApJ...748..136C,2016PhRvD..93b4013S,2017PhRvD..95d2001M,2018PhRvD..97l3014C,2018PhRvD..97f4031Z,2020ApJ...905L..25S,2021ApJ...910L..21M,2021ApJ...917L..27N,2020ApJ...902L..29N,2022arXiv220211048B,2023A&A...678A.126B,2023ApJ...959...76C,2023ApJ...958L..43H,2023arXiv230915808M}. 

Multi-messenger observations of the moments just before the merger could probe several highly-debated astrophysical scenarios \citep[see ][for a recent review, and references therein]{2021Galax...9..104W}. From a theoretical perspective, models predict the possible existence of pre-merger electromagnetic signatures via a variety of mechanisms including two-body electromagnetic interactions, resonant NS crust shattering, magnetic reconnection and particle acceleration through the revival of pulsar-like emission during the in-spiral phase, the decay of tidal tails, the formation of fireballs or wind-driven shocks \citep[e.g.,][]{1969ApJ...156...59G,1996ApJ...471L..95V,2001MNRAS.322..695H,2006MNRAS.368.1110M,2011ApJ...736L..21R,2012ApJ...757L...3L,2012ApJ...746...48M,2012ApJ...755...80P,2012PhRvL.108a1102T,2013MNRAS.431.2737M,2016MNRAS.461.4435M,2019MNRAS.488.5887S,2021ApJ...921...92B,2021MNRAS.501.3184S,2023MNRAS.519.3923C,2023PhRvL.130x5201M}. It has also been suggested that in the late in-spiral phase of a NS-NS or BH-NS merger in which one NS is a magnetar, the tidal-induced deformation may surpass the maximum that the magnetar's crust can sustain, driving a catastrophic global crust destruction that releases magnetic energy as a superflare with energy hundreds of times larger than giant flares of magnetars \citep{2022ApJ...939L..25Z}. Numerical studies support the conclusion that electromagnetic flares may be observed before the merger \citep{2013PhRvL.111f1105P,2020ApJ...893L...6M,2022MNRAS.515.2710M,2023ApJ...956L..33M}. A key related open questions is whether NS mergers may power a fraction of fast radio bursts \citep[FRBs;][]{2007Sci...318..777L,2013Sci...341...53T,2014ApJ...780L..21Z,2016ApJ...821L..22W,2019PhRvD.100d3001P,2019MNRAS.490.3483R,2020ApJ...890L..24Z,2020PTEP.2020j3E01W,2023ApJ...953..108C,2023PhRvD.108f3014P}.

Observationally, while high-energy precursors have been observed in short (and long) GRBs \citep{2005MNRAS.357..722L,2010ApJ...723.1711T,2019ApJ...884...25Z,2020ApJ...902L..42W,2022A&ARv..30....2P,2023ApJ...954L..29D}, it is still a matter of debate whether these precursors have a different origin from that of the GRB itself, or are rather just a manifestation of the variable GRB emission \citep{2015MNRAS.448.2624C,2022ApJ...941..166X}.  Searches for electromagnetic precursors have been carried in coincidence with compact binary mergers identified by LIGO and Virgo during O2/O3 having a non-negligible probability to contain a NS \citep{2022ApJ...930...45S}. While these searches  found no significant candidate precursor signals, open questions discussed above can be explored in future searches with improved sensitivity, potentially aided by GW early alerts and localizations, and extending across the electromagnetic spectrum (from radio to $\gamma$-rays; Figure \ref{fig:precursor}).

\begin{figure}
\begin{center}
\vbox{\centering
\includegraphics[width=0.51\textwidth]{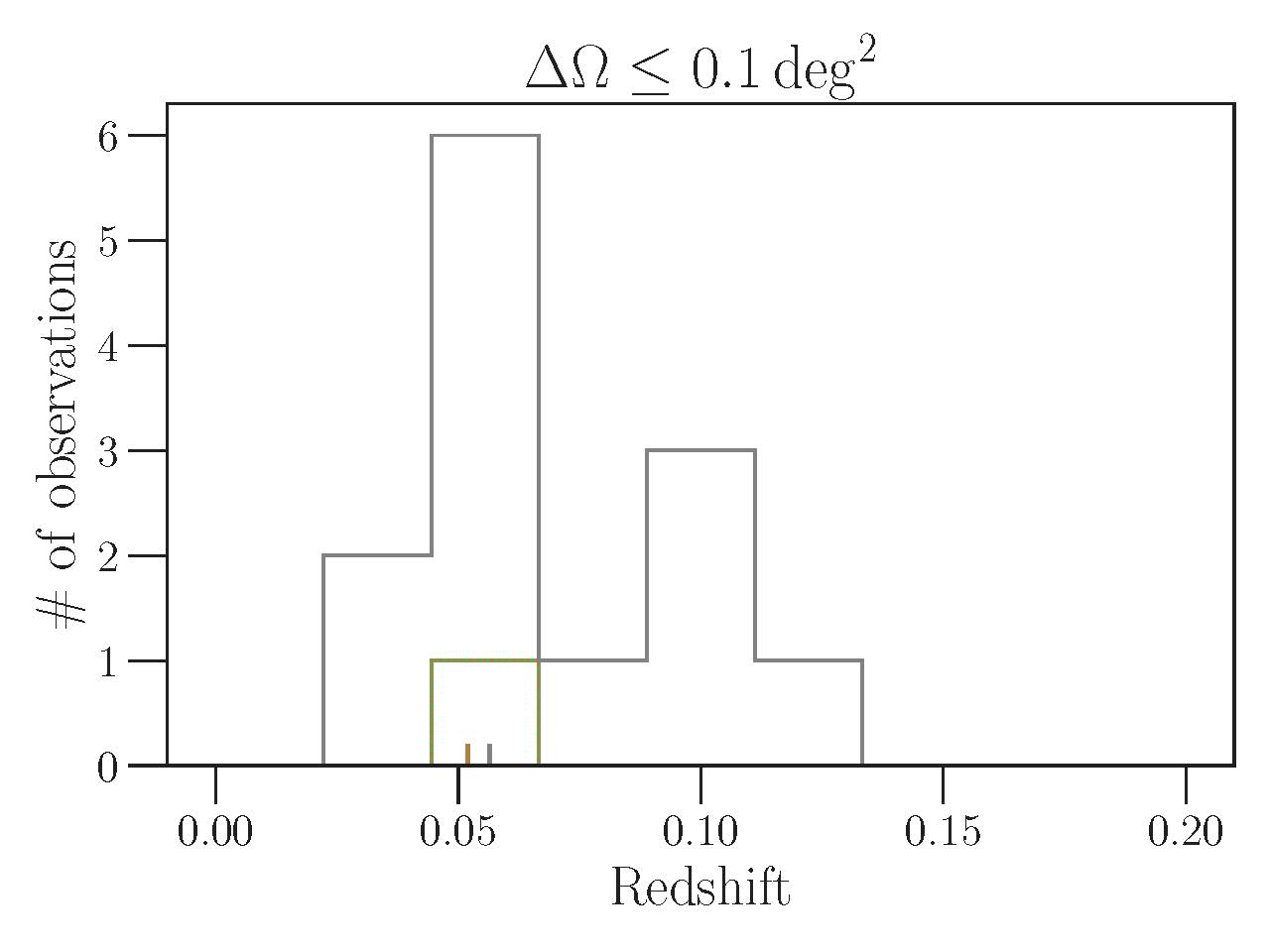}
\includegraphics[width=0.55\textwidth]{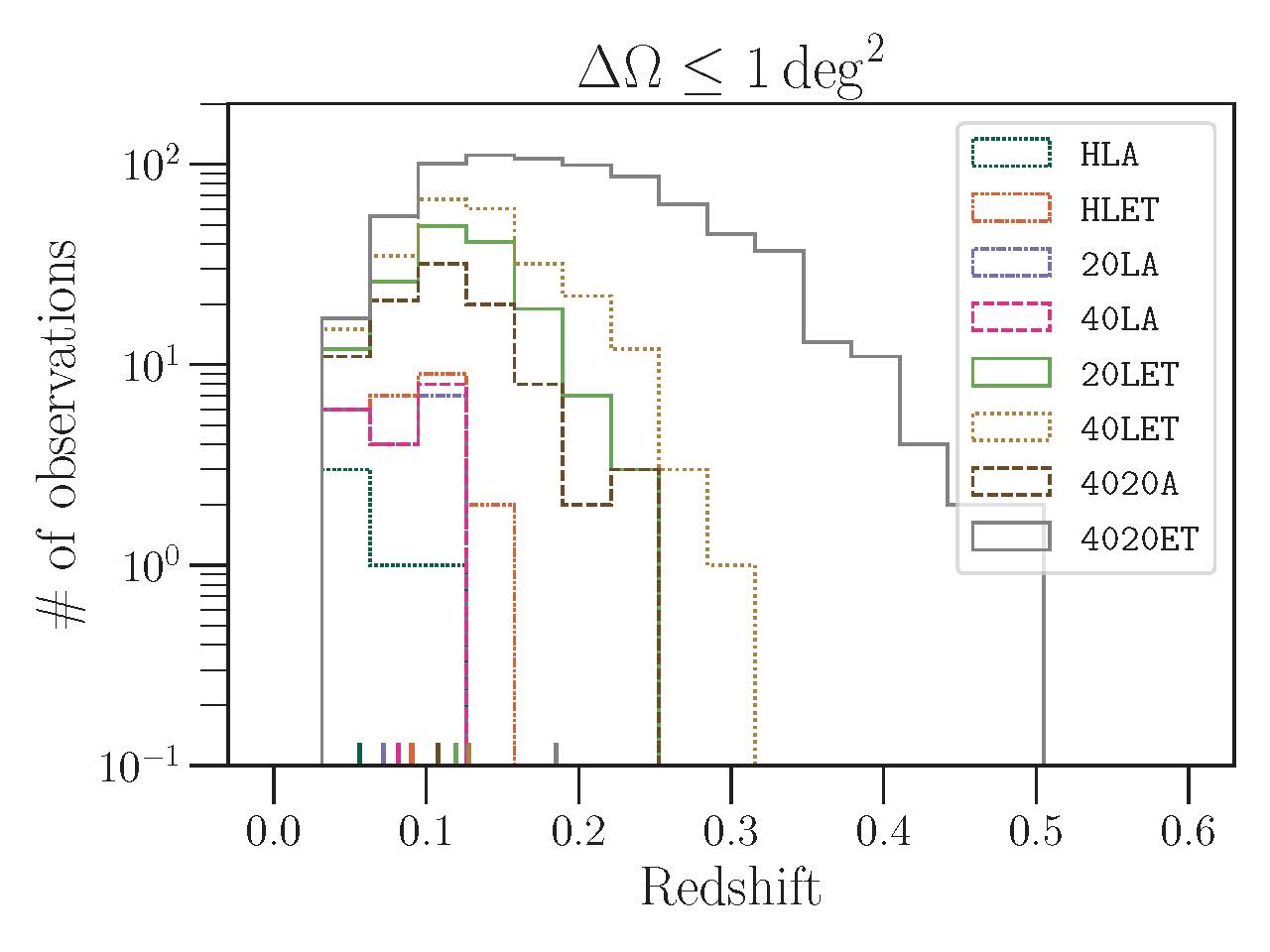}
\includegraphics[width=0.55\textwidth]{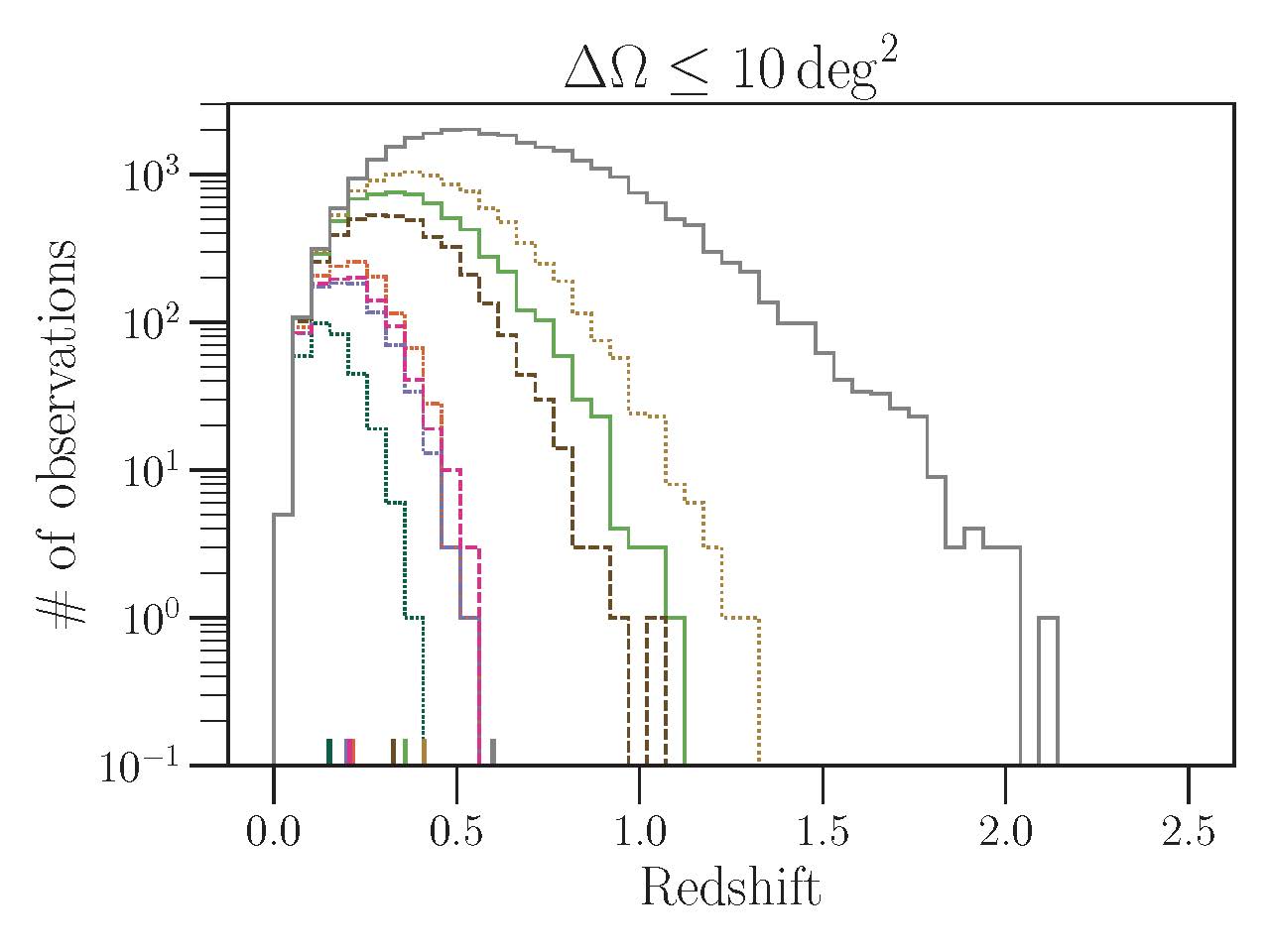}}
\caption{Figure derived from the simulations presented in \citet{2023arXiv230710421G}. Redshift distribution of NS-NS mergers detected in 1 yr and localized within the sky area indicated at the top, for various networks of ground-based GW detectors (see the caption of Figure \ref{fig:precursor}). The small vertical lines on the x-axis mark the median redshift of each distribution. The assumed local merger rate density of NS-NS systems is $320$\,Gpc$^{-3}$\,yr$^{-1}$. We note that this rate is subject to large uncertainties \citep[$10-1700$\,Gpc$^{-3}$\,yr$^{-1}$;][]{2023PhRvX..13a1048A}.}
\label{fig:redshift}
\end{center}
\end{figure}

\subsection{Nature of the merger remnant and neutron star EoS}
\label{sec:postmerger}
After a NS-NS merger, a compact remnant is left over. The nature of such a remnant---either a NS or a BH---is thought to depend primarily on the masses of the binary components (i.e., total mass of the system and mass ratio) and on the EoS of nuclear matter \citep[e.g.,][]{2014MNRAS.441.2433R,2017ApJ...844L..19P,2019ARNPS..69...41S}. If a NS remnant is formed (as opposed to a prompt BH formation), its lifetime could range from  short lived (hypermassive NS supported only temporarily against gravity by differential rotation), to long lived (supramassive NSs supported against gravity by uniform rotation), to indefinitely stable \citep[][]{2021ApJ...920..109B,2022ApJ...939...51M,2024MNRAS.527.5166W}. GWs can be used to probe the post-merger remnant via a variety of yet-to-be detected signals and, when paired with electromagnetic observations, can greatly help us understand the astrophysics of the post-merger phase. 

GWs produced by oscillations of the hot, extremely dense remnant may come into reach with improved ground-based detectors \citep[e.g.,][]{2012PhRvD..86f3001B,2014PhRvD..90f2004C,2015PhRvD..91l4056B,2016CQGra..33h5003C,2023CQGra..40u5008K}. The formation of a hypermassive NS is expected to give off quasi-periodic GWs of frequencies $\approx 2-4$\,kHz, while GWs from quasi-normal modes of promptly-formed BHs are found at higher frequencies of $\approx 6.5-7$\,kHz \citep[][]{2006PhRvD..73f4027S,2022PhRvL.128p1102B}. Hence, post-merger GW observations can be used to constrain the yet-uncertain EoS of NS matter in a way complementary to measurements of the tidal deformation of the NSs during the late in-spiral phase \citep[e.g.,][]{2008PhRvD..77b1502F,2017PhRvD..96l4035C,LandryLett}.
Simulations indicate that oscillations of a deformed, differentially rotating massive NS emit a GW spectrum with a pronounced peak generated by the fundamental quadrupolar oscillation mode, whose frequency correlates with the radius of the non-rotating NS  \citep{2012PhRvD..86f3001B}. More specifically, the frequency of
this mode is proportional to the square root
of the mean density \citep{2012PhRvD..86f3001B}. Hence, for a given remnant mass (approximately given by the total binary mass), the peak frequency is determined by the radius. In turn, the determination of the dominant post-merger GW frequency can provide an upper-limit for the maximum mass of non-rotating NSs, with implications for the NS mass distribution and, indirectly, electromagnetic counterparts \citep{2017ApJ...850L..19M,2020GReGr..52..108B,2020ApJ...893..146A}. It has also been suggested that in compact binary mergers where short-lived NSs are formed after the merger, the quasi-periodic oscillations of the remnants may imprint quasi-periodic modulations of the $\gamma$-rays emitted in the associated GRBs \citep{2023Natur.613..253C}. As of today, the viability of this process remains debated \citep{2023ApJ...947L..15M}.

After the early (dynamical) GW-driven phase, the (secular) evolution of remnants that did not collapse to BHs is driven driven by viscous magnetohydrodynamics processes and neutrino cooling \citep{2017ApJ...844L..19P,2020GReGr..52..108B}. Mapping observationally NS-NS progenitors to their remnants via their GW and electromagnetic emission offers an unprecedented opportunity to understand this complex interplay of gravitational, nuclear, weak and electromagnetic interactions \citep{2021ApJ...920..109B,2022ApJ...939...51M,2024MNRAS.527.5166W}. In the case of GW170817, the presence of an electromagnetic  counterpart disfavors a prompt BH formation. The velocity, total mass, and electron fraction of the blue kilonova ejecta (as constrained from the observations) support the idea that the merger formed a rapidly spinning hypermassive and magnetized NS, with a 0.1--1\,s lifetime \citep{2018ApJ...856..101M}.  In this interpretation, the lifetime of the GW170817 merger remnant is short because a long-lived remnant would have injected a rotational energy of a few $\approx 10^{52}$\,erg into the ejecta, which can be excluded from observations \citep{2020ARNPS..70...95R}. However, an interpretation of GW170817 in the context of a long-lived (days to months) remnant with a small dipole magnetic field (so as to minimize the energy injected into its outflows) cannot be excluded \citep{2018ApJ...860...57A,2018ApJ...861..114Y}. 

Overall, post-merger scenarios involving long-lived or stable NSs formed in compact binary mergers have been proposed to explain various features in GRB light curves and have received new attention after GW170817. Proposed electromagnetic signatures of long-lived remnants range from  brighter-than-normal magnetar-powered kilonovae, to early-time X-ray afterglow plateaus and late-time radio and X-ray flares \citep{2011Natur.478...82N,2013MNRAS.430.1061R,2018ApJ...867...95H,2019MNRAS.485.4150B,2019MNRAS.487.3914K,2021MNRAS.506.5908N,2022MNRAS.516.2614A,2022MNRAS.516.4949S,2024MNRAS.527.5166W}. Proposed GW signatures include oscillation modes of a short-lived hypermassive
NS, bar-mode instabilities, and rapid spindown powered by
magnetic-field induced ellipticities \citep[e.g.,][]{1995ApJ...442..259L,1998PhRvD..58h4020O,2002PhRvD..66h4025C,2005PhRvL..94t1101S,2009ApJ...702.1171C,2013PhRvD..88d4026H,2013MNRAS.435L..43C,2015ApJ...798...25D,2016CQGra..33h5003C}. Several observing campaigns aimed at identifying electromagnetic or GW signatures of long-lived remnants have been conducted for both GW170817 and other short GRBs, and promise to become more constraining of proposed models with next generation GW and electromagnetic instrumentation \citep[e.g.,][]{2016PhRvD..93j4059C,2016ApJ...819L..22H,2017ApJ...851L..16A,2019ApJ...875..160A,2019PhRvD.100l4041S,2020ApJ...902...82S,2021PhRvD.104j2001A,2021ApJ...914L..20B,2021MNRAS.505L..41B,2021ApJ...908...63G,2022ApJ...938...12B,2022ApJ...927L..17H,2022MNRAS.510.1902T,2023ApJ...948..125E,2024MNRAS.527.8068G,2023PhRvD.108l3045G,2023CQGra..40u5008K}. By probing the mass of the post-merger remnants in a systematic fashion, next generation GW detectors like CE and ET could also probe models of supernova engines \citep{FryerLett}.

\subsection{Compact binary merger population properties}
\label{sec:populations}
As the number of NS-NS, BH-NS, and BH-BH detections increases following the improvement in sensitivity of the LIGO, Virgo, and KAGRA detectors (Figures \ref{fig:sensitivity} and \ref{fig:redshift}), MMA studies based on single-event analyses will be crucially complemented by statistical studies of larger source samples. While interesting individual events and outliers will enable probing the most extreme systems, joint analyses of a large number of compact binaries will yield an exquisite characterization of the properties of the bulk of the population. 
 These analyses can constrain key population properties such as merger rates, mass distributions, r-process yields, properties of the GRB jets, etc. \citep[e.g.,][]{2023PhRvX..13a1048A,2023RNAAS...7..136B,2021ApJ...920L...3C,2020ApJ...893...38B,2023PhRvD.108d3023D}, while enabling comparison with similar constraints derived from observations via other messengers \citep[e.g.,][]{2021arXiv211109401B,2021ApJ...921L..25L,2022ApJ...929L..26F,2022LRR....25....1M,2023ApJ...946....4L}. On the longer term, the study of NS-NS mergers is likely to see an even more substantial shift from single-event analyses to population inference and statistical studies. In fact, next-generation GW detectors may enable us to probe the properties of NS-NS mergers across cosmic history and galactic environments (Figure \ref{fig:reach}), measure the time delay distribution between formation and merger \citep{2019ApJ...878L..13S}, and thereby infer the history of chemical evolution in the universe even beyond the reach of electromagnetic astronomy \citep[][]{2022arXiv220610622C}. For the loudest and best-localized BH-BH binaries, the uncertainty volume will be small enough to confidently identify the host galaxy even in absence of a counterpart~\citep{2018PhRvD..98b4029V,2022arXiv220211048B}.
 The ability of GW detectors to precisely measure masses, distances and sky positions of thousands of mergers per year is key to this end \citep[][see Figure \ref{fig:redshift}]{2017PhRvD..95f4052V,2023arXiv230710421G,2023arXiv230613745E}.

Increased detection rates of compact binary mergers containing the heaviest stellar-mass BHs will also shed light on crucial open questions in stellar astrophysics, especially when combined with electromagnetic surveys. Theory predicts the existence of a gap in the BH mass distribution because of pair-instability supernova \citep[][]{1964ApJS....9..201F,1967PhRvL..18..379B,2017ApJ...836..244W}.
This mechanism should produce a dearth of BH-BH binaries with components in the mass range $\sim 50-135\,M_\odot$ \citep[e.g.,][]{2016A&A...594A..97B}. The largest uncertainty on the lower end of this ``mass gap'' comes from uncertainties on the nuclear reaction rate \ensuremath{^{12}\rm{C}\left(\alpha,\gamma\right)^{16}\!\rm{O}} \citep{2019ApJ...887...53F}.
The mass gap can be contaminated from hierarchical mergers of lower-mass BHs \citep{2017PhRvD..95l4046G,2017ApJ...840L..24F,2021NatAs...5..749G} and from other formation channels with possible characteristic electromagnetic signatures, including stellar collisions in young stellar clusters \citep{2022MNRAS.516.1072C,2023MNRAS.519.5191B}; the core collapse of rapidly rotating massive stars from progenitors with helium cores $\gtrsim 130\,M_\odot$ (``collapsars''), which could lead to long GRBs, r-process nucleosynthesis, and GWs of frequency $\sim 0.1-50$\,Hz from non-axisymmetric instabilities \citep{2022ApJ...941..100S};  super-Eddington accretion in isolated binaries \citep{2020ApJ...897..100V}; or more exotic scenarios, such as accretion onto primordial BHs \citep{2021PhRvL.126e1101D,2023arXiv231214097D}. Several astrophysical scenarios predict the possibility of mergers between BHs on the ``far side'' of the mass gap \citep{2019ApJ...883L..27M,2021MNRAS.505L..69H,2023MNRAS.524..307S}.
The observation of such mergers with next-generation GW detectors could allow us to measure the location of the upper end of the mass gap. Since the ``width'' of the mass gap is to a good approximation constant as a function of the uncertain nuclear reaction rates \citep[e.g.,][]{2019ApJ...887...53F}, these constraints will also inform us about the location of the lower end of the mass gap. Theoretical predictions should also be compared with the already evident ``bump'' in the observed mass distribution of BH-BH mergers at $\sim 35\,M_\odot$ that cannot be explained by Poisson noise alone \citep{2023ApJ...955..107F}. Ultimately, the combination of GW observations and electromagnetic transient surveys can give important insight into nuclear reaction rates and supernova physics \citep{2020ApJ...902L..36F,2023MNRAS.523.4539K}.

\subsection{Impact of GW-enabled MMA on Cosmology}
\label{sec:cosmology}

Observations of GWs from well-localized compact binary mergers can measure absolute source distances. When coupled with an independent determination of redshift through an electromagnetic counterpart, they provide constraints on the Hubble constant ($H_0$) and hence the expansion history of the universe \citep[e.g.,][]{1986Natur.323..310S,2006PhRvD..74f3006D,2010ApJ...725..496N,2005ApJ...629...15H,2009LRR....12....2S,2017Natur.551...85A,2021A&A...646A..65M,JinLett,2023chep.confE.127M,2024arXiv240203120C}. Absolute distance measurements at low redshifts, as those enabled by GW observations, can  constrain dark
energy when combined with observations of the primary anisotropies in the cosmic microwave background \citep[e.g.,][]{2005ASPC..339..215H}. We note that, in modified theories of gravity that predict a non-trivial dark energy equation of state and deviations from general relativity in the propagation of GWs across cosmological distances, the effect of the modified GW propagation can dominate over that of the dark energy equation of state, potentially becoming observable with next generation GW observatories \citep[e.g.,][]{2018PhRvD..98b3510B,2021MNRAS.502.1136M,2023arXiv231216292A}.

Multi-messenger observations of GW170817 allowed for a measurement of the Hubble constant using the GW detection of the NS-NS merger combined with the optical identification of the host galaxy \citep{2017Natur.551...85A}. The GW measurement returned a value of $H_0 = 70^{+12}_{-8}$\,km\,s$^{-1}$\,Mpc$^{-1}$. While this measurement is not sufficiently precise to significantly impact the current debate on the tension between different measurements of $H_0$ \citep{2021ApJ...919...16F,2023ARNPS..73..153K,2023JCAP...11..050F}, its importance as a measurement completely independent of both the Planck cosmic microwave background and the local Cepheid-supernovae distance ladder measurements has been widely recognized. 
The dominant source of uncertainty in the $H_0$ measurement via GWs is the degeneracy between the binary viewing angle and the source distance. Hence, an independent determination of the viewing angle is of great importance \citep{2021ApJ...909..114N}. For this reason, and as demonstrated by GW170817 itself, VLBI observations of the afterglow radio centroids and images of compact binary mergers are key to improve the $H_0$ measurement \citep{2018Natur.561..355M,2019Sci...363..968G,2019NatAs...3..940H,2023MNRAS.524..403G,2023arXiv230710402C}. \cite{2019NatAs...3..940H} estimate that 15 more localized GW170817-like events (with comparable signal-to-noise ratio and favorable orientation), having radio images and light curve data, can resolve the current Hubble tension, as compared to 50-100 GW events necessary in the absence of radio data. An accurate measurement of the Hubble constant from standard siren GW cosmology also requires a robust peculiar velocity correction of the redshift of the host galaxy. The variation in the precision of $H_0$ inferred from 10 bright siren events as related to the variation in the peculiar velocity of the host galaxies can range from $\approx 5.4-6\%$ to $\approx 1.1-2.7\%$, respectively, for the LIGO-Virgo-KAGRA to CE+ET networks \citep{2024MNRAS.527.2152N}.  

It is important to note that a substantial fraction of sources detected by a given GW network over a certain timescale may not have associated transient electromagnetic counterparts. However, multi-messenger studies can still be relevant as they provide advantages related to incorporating host galaxy information. Indeed, it is
possible to carry out a measurement of $H_0$ using a statistical approach that incorporates the redshifts of all potential host galaxies within the GW three-dimensional
localization region \citep{2018Natur.562..545C}. This technique yields an $H_0$ measurement that has a greater uncertainty than that which can be achieved via direct counterpart identifications, but still informative once many detections are combined \citep{2018Natur.562..545C}. The statistical approach also implies that, in the absence of a counterpart, only those GW events with small enough localization volumes yield
informative $H_0$ measurements. Another proposed statistical technique exploits the clustering scale of the GW sources with galaxies of known redshift, and will be applicable also to the high redshift GW sources detectable with next generation GW detectors \citep{2021PhRvD.103d3520M,2022MNRAS.511.2782C,2022arXiv220303643M}. In summary, with GW detectors of improved sensitivity able to observe farther and to localize better, galaxy surveys and statistical approaches for the measurement of $H_0$ are likely to become more and more relevant \citep{2021PhRvD.104d3507Y,2023arXiv231205302B,2023arXiv231216305G,2023arXiv231008991D}. In the era of next-generation GW detectors, other statistical techniques that do not require host galaxy information nor electromagnetic counterpart identifications may  complement the constraints on cosmology as determined via these MMA techniques, particularly for the population of BH-NS mergers \citep{2023arXiv231016894C,2023ApJ...955..149S}.

\section{New Frontiers in MMA}
\label{sec:frontiers}
NSs and stellar-mass BHs, in isolation, in binary systems, and/or overall as populations, can be sources of GW signals that are very different
from the compact binary merger signals already detected by LIGO and Virgo. We have mentioned some of these signals in the context of the nature of the post-merger remnant question left open by GW170817 (Section \ref{sec:postmerger}). Here, we expand our discussion to a zoo of yet-to-be-detected signals that may reveal the physics behind a suite of extreme astrophysical phenomena, and open new ways of doing MMA that include inference of population properties via correlations between the GW signals and other (electromagnetic) observables such as galaxy counts and the cosmic microwave background \citep[e.g.,  ][]{2013RvMP...85.1401A,2020MNRAS.494.1956M,2020PhRvD.101j3509M,2022PhRvD.106d3019A,2022MNRAS.513.1105D,2023JCAP...06..050B,2023arXiv231005823D,2023PhRvD.107j2001D,2023JCAP...10..014P,2023PhRvD.108d3025Y}. 

Rotating NSs are thought to produce quasi-periodic GWs that can last for millions of years (and hence are usually referred to as continuous GWs), arising from time-varying mass quadrupoles supported by elastic or magnetic stresses \citep{Melosh1969}, or current quadrupoles known as ``$r$-modes'' \citep{Andersson_1998,Lindblom1998,2018ASSL..457..673G}. Accreting NSs (low-mass X-ray binaries), which are thought to become millisecond pulsars after accretion ends, can also be driven to non-axisymmetry by lateral temperature gradients \citep{1998ApJ...501L..89B, Ushomirsky_2002}, internal magnetic distortion \citep{Melosh1969,1996A&A...312..675B, PhysRevD.66.084025}, or magnetic bottling of accreted material \citep{2005ApJ...623.1044M}, hence emitting GWs. Continuous GW emission will help reveal properties of NSs such as composition (EoS), internal magnetic field, and viscosity, in addition to unveiling NSs that cannot be observed electromagnetically \citep[e.g.,][and references therein]{1996A&A...312..675B,1998ApJ...501L..89B,1998PhRvD..58h4020O,2001IJMPD..10..381A,2005PhRvL..95u1101O,2018ASSL..457..673G,2021MNRAS.500.5570G,2022MNRAS.517.5610M,2023LRR....26....3R}. 
 Current searches for continuous GWs produced by spinning NSs with asymmetries improve with every LIGO-Virgo-KAGRA run \citep[e.g.,][]{2022PhRvD.106j2008A} and dozens of known millisecond pulsars could come into the reach of next-generation GW detectors \citep{2018ApJ...863L..40W,2023arXiv230613745E,2023arXiv230710421G}, with the potential of many more thanks to upcoming or next generation electromagnetic facilities such as the next generation Very Large Array \citep[ngVLA;][]{2020AAS...23536401M} and the Square Kilometre Array \citep{2019BAAS...51c.239K,2023arXiv230613745E,2023ApJ...952..123P,2023LRR....26....3R,2023APh...15302880W}. Detection by next generation instruments also looks promising for bright low mass X-ray binaries such as Scorpius X-1 \citep{2023arXiv230613745E,2023arXiv230710421G}.

Impulsive, energetic NS events other than binary mergers can also produce bursts of GWs. For example, magnetar $\gamma$-ray flares \citep[possibly accompanied by FRBs;][]{2022arXiv221010931T,2022arXiv221010931T,RayLett} and pulsar glitches \citep[e.g.,][]{2022ApJ...932..133A} are the targets of current searches for GW signals in LIGO-Virgo-KAGRA data.  While near-future detector upgrades could probe GW signals expected in the most optimistic scenarios \citep{2011PhRvD..83j4014C}, next generation GW observatories are likely to probe a wider range of possible GW outcomes \citep{2023arXiv230613745E}. We stress that the detection of normal modes of NSs such as so-called ``f-modes'' will measure the cold NS EoS and masses of a population different from that seen in compact binary mergers, and combined with electromagnetic observations will yield information on internal magnetic fields \citep{2023arXiv230613745E}.

Core-collapse supernovae are also thought to generate bursts of GWs from the dynamics of hot, high density matter in their central regions. Next-generation GW detectors are expected to be sensitive to supernovae within the Milky Way and its satellite galaxies \citep{2019PhRvD.100d3026S,2022Galax..10...70S,2023arXiv230613745E,2019BAAS...51c.239K,GossanLett}, while some extreme supernovae, such as collapsars with cocoons, could generate GWs that could come into reach with current generation GW detectors \citep[e.g.,][and references therein]{2022ApJ...941..100S,2020PhRvD.101h4002A,2023ApJ...951L..30G}. The detection of a core-collapse event in GWs would provide a unique channel for observing the explosion’s central engine and the (hot) EoS of the newly formed compact remnant. A nearby supernova could also provide a spectacular multi-messenger event via a coincident neutrino detection \citep[e.g.,][]{1987PhRvL..58.1494B,2017hsn..book.1575J,2022arXiv221003088C,2023ApJ...949L..12A,2023PhRvD.108h3035G}. 

Finally, a stochastic GW background can be generated by a large variety of phenomena of cosmological \citep{2018CQGra..35p3001C} and/or astrophysical origin. 
The detection of a cosmological stochastic
background would be of fundamental importance for our understanding of the early universe. While current GW detectors are not optimized for the detection of a stochastic background of cosmological origin, a fraction of the parameter space in various scenarios is compatible with a detection by future detectors \citep{2016JCAP...04..001C,2018CQGra..35p3001C,2021PhRvD.103l3541B}. 
Astrophysical backgrounds contain key information about the distribution of mass, redshift, and other properties of their corresponding source populations \citep[e.g.,][]{2020MNRAS.491.4690M,2021MNRAS.500.1666Y}. The merger rate of NS-NS mergers as estimated from current observations suggests that distant, unresolvable binary NSs create a significant astrophysical stochastic GW background \citep{2018PhRvL.120i1101A}, adding to the contribution from BH-BH  and BH-NS binaries. In addition to compact binary coalescences of BHs and NSs, rotating NSs, magnetars, and core-collapse supernovae can all contribute sub-dominant stochastic backgrounds \citep[e.g.,][]{1998PhRvD..58h4020O,1999MNRAS.303..247F,2005PhRvD..72h4001B,2006A&A...447....1R,2011RAA....11..369R,2012PhRvD..86j4007R,2022Galax..10...34R}. 
Overall, the ability to detect and subtract GW foregrounds, and to detect sub-dominant stochastic backgrounds, is critical to unveil potential new avenues for MMA using stochastic GW signals \citep[e.g.,][]{2020PhRvD.102b4051S,2020PhRvL.125x1101B,2020PhRvD.102f3009S,2021PhRvD.104f3518M,2022arXiv220901221Z,2023PhRvD.107f4048Z,2023arXiv231002517B}.




\section{Discussion}
\label{sec:discussion}
As discussed in Section \ref{sec:diversity}, going from one GW170817-like event to $\sim 10$ well-localized NS-NS mergers detected in GWs and enjoying extensive electromagnetic follow-up represents a goal of the utmost importance for the current generation of ground-based GW detectors. It is also critical that the observational resources required to carry out a systematic electromagnetic follow up of NS-NS and BH-NS systems remain available. In fact, in the case of GW170817, observations from radio to $\gamma$-rays involving space-based and ground-based detectors with field of views (FOVs) ranging from tens of square degrees to a fraction of a square degree \citep[][and references therein]{2017ApJ...848L..12A}, all proved essential to shed light on the different ejecta components \citep[from the slow neutron-rich debris powering the kilonova to the structured jet emitting from radio to X-rays; e.g.,][and references therein]{2020LRR....23....4B,2021ARA&A..59..155M}. Going forward, it is clear that the more GW detectors improve their localization capabilities, enabling deep follow-up observations across the electromagnetic spectrum with instruments of different FOVs (Figure \ref{fig:redshift}), the larger the impact of new GW detections on the field of MMA.

Improvements in sensitivity to ground-based GW detectors will enable us to reach a GW localization accuracy of $\approx 10$\,deg$^2$ \citep[matched to the field of view of the Vera C. Rubin Observatory, hence greatly simpifying the hunt for kilonoave;][and Figure \ref{fig:redshift}]{2019ApJ...873..111I,2022ApJS..258....5A,2023PhRvD.107l4007G} for hundreds to thousands of NS-NS mergers per year with median redshifts of $z_{\rm median}\approx 0.15$ for networks containing three 4\,km-long LIGO detectors at sensitivities comparable to that of the so-called A\# configuration \citep[the ultimate performance of current LIGO detectors envisioned for the post-O5 era;][]{post-O5}; $z_{\rm median} \approx 0.2$ for networks containing at least one next-generation GW detector (with sensitivity $\approx 10\times$ that of the LIGO detectors in their projected O5 configuration); and up to a $z_{\rm median}\approx 0.6$ for an international network with three next-generation GW detectors. A network of ground-based GW detectors including one (three) next-generation instrument(s) could enable localizations of tens (hundreds and up to $\sim 10^3$) of nearby NS-NS mergers per year to $\lesssim 1$\,deg$^2$ \citep[Figure \ref{fig:redshift}; see also][]{2023arXiv230613745E,2023arXiv230710421G}. This, in turn, will allow sensitive tiling  observations of the GW error regions with radio (and X-ray) telescopes (such as the ngVLA), as well as IR telescopes \citep[such as  Nancy Grace Roman Space Telescope;][]{2019APS..APRQ09006M}, independently of a previous identification of an optical counterpart via larger FOV optical telescopes.  This capability is likely to prove critical to probe the higher-mass NS-NS and BH-NS systems that may be characterized by red and dim kilonovae, but still be accompanied by (potentially off-axis) radio-to-X-ray jet afterglows \citep{2022ApJ...927..163C,2023PhRvD.107l4007G,2024APh...15502904A}.

It is fundamental to realize that the same improvement in sensitivity that enables GW detectors in a network to localize nearby compact binary mergers to exquisite accuracy (as discussed above), also enables such detectors to see farther compact binary merger events extending the reach of MMA to higher redshifts (see Sections \ref{sec:jets}, \ref{sec:populations}, \ref{sec:cosmology}, and Figure \ref{fig:redshift}), as well as to unveil new sources of GW emission (see Sections \ref{sec:diversity}, \ref{sec:postmerger}, \ref{sec:precursosrs}, and \ref{sec:frontiers}). Indeed, as evident from the maximum redshift in the distributions in Figure \ref{fig:redshift}, only networks of next-generation detectors can extend the reach of GWs to the peak of star formation ($z\approx 1-2$) for GW events localized to $\lesssim 10$\,deg$^2$. Space missions such as \textit{Fermi} and \textit{Swift}, Roman, and future NASA programs focused on the transient and time-variable
universe, are key to ensure continued progress in the electromagentic follow-up of these events \citep{NAP26141,SambrunaLett}. From the ground, the Rubin Observatory, the Extremely Large
Telescopes, and the ngVLA will provide follow-up capabilities for GW events that are key to enable MMA to reach its full potential over the next decade and beyond \citep{2019BAAS...51g..88B,2019BAAS...51c.237C,2019arXiv190310589C,2019AAS...23336125L,NAP26141,MurphyLett}. The IceCube-Generation 2 neutrino observatory will help constrain emission models for high-energy neutrinos in nearby NS-NS mergers and potentially open the way for discoveries across three different messengers \citep{2021JPhG...48f0501A,NAP26141}. Multi-band GW data sets formed with the LISA space-based GW detector can also impact MMA studies of compact binary mergers \citep[see ][and references therein]{2016PhRvL.117e1102V,2023LRR....26....2A}.





\section*{Conflict of Interest Statement}
The authors declare that the research was conducted in the absence of any commercial or financial relationships that could be construed as a potential conflict of interest.

\section*{Author Contributions}
A.Corsi led the paper organization and writing effort. 
J.R.Smith contributed to the figures and descriptions of the GW detectors. 
B.Rajbhandari added to the prospect of continuous gravitational waves. B.Owen contributed text to the Section on New Frontiers and extensively proofread the paper. 
E.Berti and K.Kritos contributed to the section on compact binary merger population properties and to the section on the impact of MMA on cosmology. I.Gupta contributed to the figures on sky localization and early warning alerts for NS-NS mergers. A.Nitz contributed to the figure on observatory reach. D.H.Shoemaker provided extensive comments on both content and presentation. L.Barsotti and M.Evans contributed to the review of text referring to the next generation GW detectors. S.B.Sathyaprakash contributed to the overall supervision of simulations that were used for the figures on sky localization and early warning. S.Vitale contributed to the Section on Compact Binary Merger Populations. K.Kuns contributed to the figure on detectors' sensitivity. J.Read contributed to reviewing the Section on the nature of the merger remnants and the NS EoS. All authors have contributed to proofreading the paper and including appropriate references. We note that the field of MMA is progressing very rapidly, hence references cited are extensive but by no means exhaustive. 

\section*{Funding}
A.Corsi acknowledges support from NSF Grants No. AST-2307358 and PHY-2011608.
E.Berti and K.Kritos are supported by NSF Grants No. AST-2006538, PHY-2207502, PHY-090003 and PHY-20043, by NASA Grants No. 20-LPS20-0011 and 21-ATP21-0010, by the John Templeton Foundation Grant 62840, by the Simons Foundation, and by the Italian Ministry of Foreign Affairs and International Cooperation Grant No.~PGR01167. 
K.Kritos is supported by the Onassis Foundation - Scholarship ID: F ZT 041-1/2023-2024.
J.R.Smith acknowledges support from the Dan Black Family Trust, Nicholas and Lee Begovich, and NSF Grants No. PHY-2308985, AST-2219109, and PHY-2207998. J. Read acknowledges support from Nicholas and Lee Begovich, AST-2219109 and PHY-2110441.  I.Gupta and B.S.Sathyaprakash acknowledge support from NSF Grants No. PHY-2207638, AST-2307147 and PHY-2308886. B.Rajbhandari is supported by the NSF Grant 2110460. A.H.Nitz is supported by the NSF Grant PHY-2309240.
B.J.Owen acknowledges NSF Grant 2309305. D.H.Shoemaker acknowledges support from NASA award 80NSSC23K1242 and from NSF Grant No. PHY-18671764464. M.Evans, L.Barsotti, B.S. Sathyaprakash, D.H.Shoemaker, and S.Vitale acknowledge support from NSF Grant No. PHY-2309064. S.Vitale also acknowledges support from NSF Grant No. PHY-2045740.

\section*{Acknowledgments}
We thank Patrick Brady for providing useful comments on this manuscript.

\bibliographystyle{Frontiers-Harvard} 
\bibliography{main-biblio}





\end{document}